\documentstyle[12pt]{article}
\catcode`@=11
\newcount\@tempcntc
\def\@citex[#1]#2{\if@filesw\immediate\write\@auxout{\string\citation{#2}}\fi
  \@tempcnta\z@\@tempcntb\m@ne\def\@citea{}\@cite{\@for\@citeb:=#2\do
    {\@ifundefined
       {b@\@citeb}{\@citeo\@tempcntb\m@ne\@citea\def\@citea{,}{\bf ?}\@warning
       {Citation `\@citeb' on page \thepage \space undefined}}%
    {\setbox\z@\hbox{\global\@tempcntc0\csname b@\@citeb\endcsname\relax}%
     \ifnum\@tempcntc=\z@ \@citeo\@tempcntb\m@ne
       \@citea\def\@citea{,}\hbox{\csname b@\@citeb\endcsname}%
     \else
      \advance\@tempcntb\@ne
      \ifnum\@tempcntb=\@tempcntc
      \else\advance\@tempcntb\m@ne\@citeo
      \@tempcnta\@tempcntc\@tempcntb\@tempcntc\fi\fi}}\@citeo}{#1}}
\def\@citeo{\ifnum\@tempcnta>\@tempcntb\else\@citea\def\@citea{,}%
  \ifnum\@tempcnta=\@tempcntb\the\@tempcnta\else
   {\advance\@tempcnta\@ne\ifnum\@tempcnta=\@tempcntb \else \def\@citea{--}\fi
    \advance\@tempcnta\m@ne\the\@tempcnta\@citea\the\@tempcntb}\fi\fi}
\catcode`@=12

\begin{document}

\begin{flushright}
MZ-TH/91-32 \\
December 1991
\end{flushright}
\bigskip

\begin{center}
{\Large\bf Radiatively Induced Neutrino Masses and} \\[0.33cm]
{\Large\bf Large Higgs-Neutrino Couplings in the} \\[0.33cm]
{\Large\bf Standard Model with Majorana Fields} \\
\vspace{1.5cm}
{\large Apostolos Pilaftsis} \\[.2cm]
 Institut f\"ur Physik \\
 Johannes-Gutenberg-Universit\"at \\
 Staudinger Weg 7, Postfach 3980 \\
 D-6500 Mainz, Germany
\end{center}

\bigskip\bigskip\bigskip
\centerline {\bf ABSTRACT} 
\noindent
The Higgs sector of the Standard Model $(SM)$ with one right-handed 
neutrino per family is systematically analyzed. 
In a model with intergenerational independent mixings between 
families, we can account for very light neutrinos acquiring Majorana 
masses radiatively at the first electroweak loop level. 
We also find that in such a 
scenario the Higgs coupling to the light--heavy neutrinos and to the 
heavy--heavy ones may be remarkably enhanced with significant implications
for the production of these heavy neutrinos at high energy colliders.
 
\bigskip\bigskip
\centerline{\em Published in Z. Phys.\ C55 (1992) 275--282}

\newpage

\setcounter{equation}{0}
\section{Introduction}
\indent

One of the outstanding problems in particle and astrophysics is 
connected with the question of the neutrino mass, which if nonzero, has to 
be very small in the range of 10~eV for cosmological reasons~[1].
Such small neutrino masses could also eventually account for most of the
dark matter in the universe~[2,3]. Also, small mass differences between 
neutrinos ($\Delta m_{\nu} \sim 10^{-2}-10^{-6}$) 
may resolve the Solar neutrino
problem~[4]. Finally, double-beta decay experiments seem also to favour 
neutrino masses $\stackrel{\textstyle<}{\sim} 10$~eV~[5]. 
In the minimal $SM$, these 
particles are taken to be massless. In other extensions of it, a desired
nonzero Majorana mass can be obtained by including right-handed neutrino 
fields, where the well known "see-saw" mechanism takes place generating
very light neutrinos~[6]. In most theories, the mass of
the light neutrinos ${\nu}_i$ is related to that of the heavy ones $N_i$
via $m_{\nu} \sim m_D^2/m_N$, where $m_D$ is the Dirac mass scale. 
Therefore, in order to naturally provide very small neutrino masses, one has 
to impose a very large scale on $m_N$ (e.g. $10^7-10^8$~GeV, for $m_D \sim
m_{leptons}$ and $m_{\nu} \le 10-40$~eV)~[7]. However, we will show that
large intergenerational indepedent mixings of the order of $0.1$ 
may give rise to $m_{\nu} \le 10$~eV at the 
first electroweak loop level, with $m_N$ being in the 100~GeV range.
These heavy neutrinos can copiously 
be produced in the forthcoming $e^+e^-$ or hadron
colliders with typical signals being like-sign dilepton pairs and 
additional jets with no missing transverse momentum $p_T$~[8].
We also find that in some scenarios the Higgs coupling to $\nu N$ and $NN$
can significantly contribute to the production of these heavy neutrinos.

This work has been organized as follows: In section~2 we give the 
description of the $SM$ in which one right-handed neutrino field for each 
family has been introduced. We carefully study the Higgs sector of the 
model and will give some constraints on the mixing parameters involved. In
section~3 we elaborate an illustrative example considering the case where 
only two neutrino species mix with each other. In particular, we present 
the main theoretical features of the mass matrices that describe two
massless neutrinos at the tree level. In sections~4 and~5 we calculate
the radiatively induced Majorana masses for the light neutrinos and 
address numerically the issue of the implications 
of large $H-\nu-N$ and $H-N-N$ couplings
for the production cross sections of the heavy neutrinos, respectively.
Finally, in section~6 we summarize our conclusions.

\setcounter{equation}{0}
\section{The Standard Model with right-handed neutrinos}
\indent

Let us start the discussion by giving a general description of the
Yukawa sector of the $SM$ with right-handed neutrinos. 
After spontaneous symmetry breaking the relevant part
of the Lagrangian containing Dirac and Majorana mass terms is given by
\begin{equation}
-L_{mass}^{\nu} = \overline{\nu}_{R_i}^0 m_{D_{ij}}^{\dag} 
\nu^0_{L_j} + \overline{\nu}^0_{L_i} m_{D_{ij}} \nu^0_{R_j} +
\frac{1}{2} \overline{\nu}^{0C}_{R_i} m_{M_{ij}}^{\dag} \nu^0_{R_j} +
\frac{1}{2} \overline{\nu}^0_{R_i} m_{M_{ij}} \nu^{0C}_{R_j} 
\end{equation}  
In eq.~(1) $m_D$ and $m_M$ are Dirac and Majorana $n_G \times n_G$ mass
matrices, respectively and $\nu^0_L (\nu^0_R)$ is the left(right)-handed 
Weyl spinor which describes the neutrino field.  Note that $L_{mass}^{\nu}$
has the most general form which is invariant under the gauge transformation
$SU(2)_L \bigotimes U(1)_Y$. In other words, in~(2.1) we assume the absence of 
isotriplet Higgs scalars and Majoron fields~[9].
Thus, we can now express $L_{mass}^{\nu}$
in terms of the Majorana fields
\begin{eqnarray}
  f &= \nu^0_L + {(\nu^0_L)}^C \nonumber\\
  F &= \nu^0_R + {(\nu^0_R)}^C
\end{eqnarray}  
as follows:
\begin{equation}  
-L_{mass}^{\nu} = \frac{1}{2}
 {(\overline{f}_L, \overline{F}_L)}_i M^{\nu}_{ij}
{ \left( \begin{array}{c} f_R \\ F_R \end{array} \right) }_j \quad + \quad
h.c.
\end{equation} 
with
\begin{equation}
M^{\nu} = \left( \begin{array}{cc}
0 & m_D \\ m^{\rm T}_D & m_M  \end{array} \right)
\end{equation}  
Applying the properties of Majorana fields to eq.~(2.3), we find that 
$M^{\nu}$ is generally a complex symmetric matrix $(M^{\nu}=M^{\nu \rm T})$.
The matrix $M^{\nu}$ can be diagonalized 
by a $2n_G \times 2n_G$ unitary matrix 
$U^{\nu}$ in the following way: 
\begin{equation}
U^{\nu T} M^{\nu} U^{\nu} =  {\hat{M}}^{\nu}
\end{equation}    
At the same time, the Majorana fields have to be transformed according to
\begin{equation}
\left( \begin{array}{c}
f_R \\ F_R \end{array} \right) = U^{\nu}
\left( \begin{array}{c}
{\nu}_R \\ N_R \end{array} \right) \quad , \quad
\left( \begin{array}{c}
f_L \\ F_L \end{array} \right) = U^{\nu \ast}
\left( \begin{array}{c}
{\nu}_L \\ N_L \end{array} \right) 
\end{equation}  
Diagonalizing the mass matrix $M^{\nu}$, we obtain $n_G$ light neutrinos
$({\nu}_i)$ and $n_G$ heavy ones $(N_i)$.  This and  other constraints on the 
structure of $m_D$, $m_M$ and neutrino mixings, which we shall give below,
are imposed by phenomenology.

We are now in the position to give explicitly the Lagrangian which describes 
the interaction between Majorana neutrinos and gauge or Higgs bosons in 
terms of mass eigenstates.
\begin{eqnarray}
L_{int}^{W-{\nu}_M-l} & = & -\frac{g_W}{2\sqrt{2}} W^{-\mu} \Big[ \ 
 {\overline{l}}_i {\gamma}_{\mu} {\gamma}_- 
B_{l_i{\nu}_j} {\nu}_j + 
{\overline{l}}_i
{\gamma}_{\mu} {\gamma}_-
B_{l_iN_j} N_j \ \Big] \quad +\quad h.c.\nonumber\\
 & &\gamma_{-} = \ 1-{\gamma}_5  \\ 
 & & \nonumber\\
L_{int}^{Z-{\nu}_M-{\nu}_M} & = & -\frac{g_W}{4\cos {\theta}_W} Z^{0 \mu}
\Big[\ {\overline{\nu}}_i {\gamma}_{\mu}[ iIm(C_{{\nu}_i{\nu}_j})
 - {\gamma}_5 Re(C_{{\nu}_i{\nu}_j})] {\nu}_j \nonumber\\
 & & + \ ({\overline{\nu}}_i {\gamma}_{\mu} [ iIm(C_{{\nu}_iN_j}) 
- {\gamma}_5 Re(C_{{\nu}_iN_j})] N_j \quad + \quad h.c.) \nonumber\\
 & & + \ {\overline{N}}_i {\gamma}_{\mu}[ iIm(C_{N_iN_j}) - {\gamma}_5
Re(C_{N_iN_j})] N_j \ \Big] \\
 & & \nonumber\\
L_{int}^{H-{\nu}_M-{\nu}_M} & = & - \frac{g_W}{4M_W} H^0
\Big[\ {\overline{\nu}}_i [(m_{{\nu}_i}+m_{{\nu}_j})Re(C_{{\nu}_i{\nu}_j})
+i{\gamma}_5(m_{{\nu}_j}-m_{{\nu}_i})Im(C_{{\nu}_i{\nu}_j})]{\nu}_j   
\nonumber\\ 
 & & + \ 2 {\overline{{\nu}}}_i [(m_{{\nu}_i}+m_{N_j})Re(C_{{\nu}_iN_j})
 + i {\gamma}_5 (m_{N_j}-m_{{\nu}_i})Im(C_{{\nu}_iN_j})] N_j 
\nonumber\\
 & & +  \ {\overline{N}}_i [(m_{N_i}+m_{N_j})Re(C_{N_iN_j})
 + i {\gamma}_5 (m_{N_j}-m_{N_i})Im(C_{N_iN_j})] N_j \ \Big] \nonumber\\
 & &
\end{eqnarray} 
The matrices $B$ and $C$ given above can be  expressed
in terms of $U^{\nu}$ by
\begin{eqnarray}
B_{l_ij} & = & \sum\limits_{k=1}^{n_G} V^l_{ik} U^{\nu\ast}_{kj}
\quad  \quad \mbox{with} \quad j=1,\dots,2n_G \\
C_{ij} & = & \sum\limits_{k=1}^{n_G} U^{\nu}_{ki} U^{\nu\ast}_{kj}
\quad \quad \mbox{with} \quad i,j=1,2,\dots,2n_G 
\end{eqnarray} 
where $V^l$ in eq.~(2.10) is the corresponding Kobayashi-Maskawa ($KM$) matrix 
for the lepton sector. For a proper labeling of the neutrino fields in
eqs~(2.10), (2.11), notice that the indices $i$, $j$ refer for  ${\nu}_i$
or ${\nu}_j$ to $i,j=1,\dots,n_G$ and for $N_i$ or $N_j$ to $i,j=
n_G+1,\dots,2n_G$.
Furthermore, the matrices $B$, $C$ obey the following equalities:
\begin{eqnarray}
\sum\limits_{k=1}^{2n_G} B_{l_ik}B^{\ast}_{l_jk} & = &{\delta}_{l_il_j}\\
\sum\limits_{k=1}^{n_G} B_{l_kj}B^{\ast}_{l_ki} & = & C_{ij}
\quad \mbox{with} \quad i,j=1,\dots,2n_G\\
\sum\limits_{k=1}^{2n_G} C_{ik}C^{\ast}_{jk} & = & C_{ij}
\quad \mbox{with} \quad i,j=1,\dots,2n_G
\end{eqnarray} 
Eq.~(2.12) can be regarded as the generalized form of the unitarity 
condition for $n_G$ charged leptons and $2n_G$ neutrinos. Since in 
section~4 we will calculate neutrino masses
induced by loop corrections in the Feynman gauge, we,
for definiteness, give the relevant couplings of the Majorana neutrinos 
with the unphysical Goldstone bosons ${\chi}^-$ and ${\chi}^0$.
\begin{eqnarray}
 & & \nonumber\\[.3cm]
L_{int}^{{\chi}^--{\nu}_M-l} & = & -\frac{g_W}{2\sqrt{2}M_W} {\chi}^- \Big[ 
\  {\overline{l}}_i [m_{l_i}B_{l_i{\nu}_j} {\gamma}_--{\gamma}_+
B_{l_i{\nu}_j}m_{{\nu}_j} ] {\nu}_j\nonumber\\
 && + \ {\overline{l}}_i [m_{l_i}B_{l_iN_j} {\gamma}_--{\gamma}_+
B_{l_iN_j}m_{N_j} ] N_j \ \Big]\quad + \quad h.c.\\
& & \nonumber\\
L_{int}^{{\chi}^0-{\nu}_M-{\nu}_M} & = & - \frac{ig_W}{4M_W} {\chi}^0
\Big[\ {\overline{\nu}}_i [(m_{{\nu}_i}+m_{{\nu}_j}){\gamma}_5
Re(C_{{\nu}_i{\nu}_j})+i(m_{{\nu}_j}-m_{{\nu}_i})Im(C_{{\nu}_i{\nu}_j})]
{\nu}_j \nonumber\\ 
 & & + \ 2 {\overline{{\nu}}}_i [(m_{{\nu}_i}+m_{N_j})
{\gamma}_5Re(C_{{\nu}_iN_j})
 + i (m_{N_j}-m_{{\nu}_i})Im(C_{{\nu}_iN_j})] N_j
\nonumber\\
 & & +  \ {\overline{N}}_i [(m_{N_i}+m_{N_j}){\gamma}_5Re(C_{N_iN_j})
 + i(m_{N_j}-m_{N_i})Im(C_{N_iN_j})] N_j  \ \Big] \nonumber\\
 & &
\end{eqnarray} 
 
Up to now we have treated the neutrinos as nonzero mass particles and 
therefore our Lagrangians derived above are of general use. One can now 
easily prove the following theorem:\\
{\em A sufficient and necessary condition for the j-th light 
neutrino to be massless already at tree level is}
\begin{equation}
\sum\limits_{k=1}^{2n_G} M^{\nu}_{ik} U^{\nu}_{kj} = 0 
\quad \mbox{for each} \quad i=1,2,\dots,2n_G
\end{equation} 
Assuming now all light neutrinos ${\nu}_i$ to be massless, one finds that 
couplings proportional to $m_{{\nu}_i}$ disappear in the Lagrangians
(2.9), (2.15) and (2.16). Nevertheless, what kind of structure of 
$M^{\nu}$ follows from restriction (2.17), will be seen in the context
of a two generation model $(n_G=2)$ in section~3.

Sometimes, in order to avoid excessive complication in our calculations,
we expand $U^{\nu}$ in power series of the matrix 
parameter ${\xi} = m_D m_M^{-1}$ [10], with the constraint
${\xi}_{ij} < 1$.  The form of $U^{\nu}$ to the third order of ${\xi}$ can 
be estimated to be
\begin{equation}
U^{\nu} = \left(
\begin{array}{cc}
1-\frac{1}{2} {\xi}^{\ast}{\xi}^T & {\xi}^{\ast}(1-\frac{1}{2} {\xi}^T
{\xi}^{\ast})J \\
-{\xi}^T(1-\frac{1}{2} {\xi}^{\ast}{\xi}^T) & 
(1-\frac{1}{2} {\xi}^T{\xi}^{\ast})J
\end{array}  \right)
\qquad + \qquad {\cal O}({\xi}^4)
\end{equation}   
where $J$ is a $n_G \times n_G$ diagonal unitary matrix, which 
guarantees that the nonzero mass eigenvalues have positive values.
Up to next to leading order in ${\xi}$ the neutrino masses are given
by
\begin{eqnarray}
m_{\nu} & = & m_D {\xi}^T = {\xi}m^T_D = 0 \\
m_N & = & J m_M[ \ 1+\frac{1}{2m_M}({\xi}^{\dagger}m_D
+m_D^T{\xi}^{\ast}) + {\cal O}({\xi}^3) \ ]J 
\end{eqnarray} 
Making also use of eq.~(2.18) we find that the matrices $B$, $C$ in this
approximation are determined by
\begin{eqnarray}
B_{l_i{\nu}_j}  =  [V^l(1-\frac{1}{2}\xi{\xi}^{\dagger})]_{l_i{\nu}_j}
\quad, \quad B_{l_iN_j}=
[V^l\xi(1-\frac{1}{2}{\xi}^{\dagger}\xi)J^{\ast}]_{l_iN_j}\\
C_{{\nu}_i{\nu}_j} = (1-\xi{\xi}^{\dagger})_{{\nu}_i{\nu}_j} \ , \
\ C_{{\nu}_iN_j} = [\xi(1-{\xi}^{\dagger}\xi)J^{\ast}]_{{\nu}_iN_j} \ ,
\ \ C_{N_iN_j} = [J{\xi}^{\dagger}\xi J^{\ast}]_{N_iN_j}
\end{eqnarray} 
The mixing couplings $C_{{\nu}_i{\nu}_j}$ can be taken to be diagonal, 
because the masslessness condition~(2.17) gives the freedom to rotate 
the light neutrino fields by an arbitrary unitary matrix $R$ as
\begin{equation}
{\nu'}_i \ = \ R_{ij}{\nu}_j
\end{equation} 
The matrix $C$ spanned in the massless neutrino space is hermitian and 
can hence be diagonalized by choosing the unitary matrix
$R$ appropriately.

The allowed values of the mixing parameters ${\xi}_{ij}$ have been 
systematically investigated in [11], where a global analysis based on
charge-current universality, neutral-current effects and other experimental
constraints has been performed. In fact, it has been shown that these 
parameters have maximal values of the order of $0.1-0.2$, where the larger
bound applies safely to the systems $e-\tau$, $\mu - \tau$~[12]. Such 
large mixings are even preferred by some neutrino-mass schemes~[13] for
resolving the solar-neutrino problem through the $MSW$ mechanism~[4].
On the other hand, heavy neutrinos $N_i$ with mixings ${\xi}_{ij}>3$~$  
10^{-2}$ must be heavier than the $Z^0$ boson, since otherwise they could
already be produced in $Z^0$ decays at $LEP$.

For the $H^0-\nu-N$ and $H^0-N-N$ couplings some comments are in order
here. These interactions are significantly enhanced for heavy neutrinos
with $m_N \gg M_W$. From eqs~(2.8) and~(2.9) we remark that the $H^0-
\nu-N$ coupling is almost by a factor $m_N/M_W$ larger than the $Z^0-
\nu-N$ coupling, while the $H^0-N-N$ coupling is $2m_N/M_W$ times
as large as the $Z^0$ corresponding one (i.e. $Z^0-N-N$). This fact
may have important implications for the production cross sections of these
heavy neutral leptons at high energies~[14]. We will address this issue
numerically in section~5. Finally, it should be also noted that the 
Dirac mass terms $m_{D_{ij}}$ defined in~(2.1) cannot be arbitrarily
large, since they are constrained by renormalization-group-triviality
bounds ( $m_{D_{ij}} \stackrel{\textstyle<}{\sim} 0.3$~TeV)~[15].

\setcounter{equation}{0}
\section{The neutrino mass matrix -- the case $n_G = 2$}
\indent

For the sake of illustration, we will now give the main theoretical 
characteristics for the mass matrices describing two massless neutrinos 
only ($n_G=2$). The general form of the matrices $m_D$ and $m_M$ 
as parameterized in~[16] is given by
\begin{equation}
m_D = \left( 
\begin{array}{cc}
a & b \\
c & d    \end{array} \right) \qquad \qquad \qquad
m_M = \left( \begin{array}{cc}
A & 0 \\
0 & B    \end{array} \right)
\end{equation} 
where $m_D$ is a general complex matrix and $m_M$ can be chosen to be
real.  Now, the requirement for two zero eigenvalues corresponding to
the two massless neutrinos prescribes that $M^{\nu}$ fulfills the
following two conditions:
\begin{eqnarray}
\prod\limits_{i=1}^4 m^2_i & = & \det (M^{\nu}M^{\nu\dagger}) = 0 \\
\sum\limits_{i<j<k}^4 m^2_i m^2_j m^2_k & = & \frac{1}{6}
\Big[ tr^3(M^{\nu}M^{\nu\dagger})
 + 2tr(M^{\nu}M^{\nu\dagger})^3 
- 3tr(M^{\nu}M^{\nu\dagger})tr(M^{\nu}M^{\nu\dagger})^2 \Big]
 = 0 \nonumber\\
 & & 
\end{eqnarray} 
Equation~(3.2) leads automatically to the constraint: 
\begin{equation}
\mbox{det\hspace{0.08cm}}m_D \ = \ 0
\end{equation} 
In particular, assuming without any loss of generality that $a \not= 0$, we
can derive from eqs~(3.3) and~(3.4) that
\begin{equation}
d = \frac{bc}{a} \quad, \qquad B = - \frac{b^2}{a^2} A
\end{equation} 
We also find that the parameters $a$, $b$ ,$c$ ,$d$  should be either purely
real or purely imaginary numbers.
Taking, for example, these parameters to be real, we can evaluate the 
masses of the heavy neutrinos to be 
\begin{eqnarray}
m_{N_1} & = & \frac{A}{2} \Big[ 1 - \frac{b^2}{a^2} + \left(
1 + \frac{b^2}{a^2} \right) \sqrt{1 + \frac{4a^2}{a^2+b^2} \left(
1 + \frac{c^2}{a^2} \right) \frac{a^2}{A^2} } \Big] = A + {\cal O}(1/A) 
\nonumber\\
m_{N_2} & = & -\frac{A}{2} \Big[ 1 - \frac{b^2}{a^2} - \left(
1 + \frac{b^2}{a^2} \right) \sqrt{1 + \frac{4a^2}{a^2+b^2} \left(
1 + \frac{c^2}{a^2} \right) \frac{a^2}{A^2} } \Big] = 
\frac{b^2}{a^2} A + {\cal O}(1/A) \nonumber\\
 & &  
\end{eqnarray} 
In this scheme the matrix $J$ turns out to be
\begin{equation}
J = \left( \begin{array}{cc}
1 & 0 \\
0 & i   \end{array} \right)
\end{equation}  
Notice that even in case $a=b=c$, we have $d=a$ and $B=-A$, which
means that models with family-independent mixings can naturally account
for massless neutrinos already at the tree level. In reality one has
to assume that the Dirac mass terms are described by a universal
Yukawa coupling $a$ and the two right-handed weak eigenstates
${\nu}^0_{R_{1,2}}$ possess opposite $CP$~quantum numbers. However, in
our forthcoming calculations we will consider that there exist small
perturbations on this family-independent scenario in such a way that
eq.~(3.5) is always valid. Recently, patterns with "democratic"-type
mixing between quark families have also been proposed in~[17] in order
to explain the structure of the usual $KM$-mixing matrix.

So far,  the "see-saw" mechanism is the only 
known scheme~[6,7] for naturally generating small, nonvanishing, neutrino
masses. According to this mechanism,  the Majorana mass terms 
($m_M \sim A \sim m_N$) in eq.~(3.1) can be regarded as remnant parts
of a more fundamental theory, which contains the $SM$ as an effective
low energy approximation. Some examples could be the Left--Right symmetric
models~[18] or grand unified models ($GUT$),
 e.g. $SO(10)$~[19], or more complicated
patterns arising from certain embeddings into the gauge group $E_6$ [19]. 
In all these models it is possible to have TeV-mass scales determined by 
the breaking mechanism itself, as the symmetry of the original gauge
group breaks spontaneously down to $SU(2)_L \bigotimes U(1)_Y$.
In some $GUT$~theories one also obtains 
that the Dirac mass matrix of the neutrinos
equals, up to a factor of proportionality of the order one,
the quark or the charged-lepton mass matrix [6,19]. Thus, in order
to have $m_{{\nu}_i} \stackrel{\textstyle<}{\sim} 1-10$~eV
for $m_D \approx 1$~GeV through the "see-saw" mechanism, one must
require that  very heavy neutrinos 
with masses $m_{N_i} \stackrel{\textstyle>}{\sim}
10^7-10^8$~GeV are present. However, as we will see in the next
section, in the context of the model outlined above small
neutrino masses $m_{\nu}$ can naturally be induced
by radiative corrections with $m_{N_i} \sim 100$~GeV and 
${\xi}_{ij} \sim 0.1$.

\setcounter{equation}{0}
\section{Radiatively induced neutrino masses}
\indent

We will now proceed with the calculation of the Feynman graphs depicted
in fig.~1, which give rise to one-loop neutrino masses of the Majorana type.
Note that only the heavy--light neutrino couplings with the Higgs and 
$Z^0$ boson are responsible for the generation of a nonzero neutrino
mass matrix given by 
\begin{equation}
m^{\nu}_{{\nu}_i{\nu}_j} \ = \ {\Sigma}^{{\nu}_i{\nu}_j}(\not\! q)
 \Big|_{{\not\! q} = 0}
\end{equation} 
where ${\Sigma}^{{\nu}_i{\nu}_j}(\not\!\! q)$ is the selfenergy diagram of 
neutrinos. In particular, working in the on-shell renormalization 
scheme and adopting the Feynman gauge~[20], we find that the graphs~(1d)-(1f)
do not contribute to $m^{\nu}$. The graph~(1b) is finite by itself, 
while the ultraviolet divergences existing in~(1a) and~(1c) cancel 
each other. Thus, we finally arrive at the following expression for the 
neutrino mass matrix:
\begin{equation}
m^{\nu}_{{\nu}_i{\nu}_j} \ = \ \frac{{\alpha}_W}{16\pi}
\sum\limits_{k=1}^{n_G} \frac{m_{N_k}}{M^2_W} C_{{\nu}_iN_k}
C_{{\nu}_jN_k} F(m^2_{N_k},M^2_Z,M^2_H)
\end{equation} 
with
\begin{equation}
F(m^2_{N_k},M^2_Z,M^2_H) \ = \ m^2_{N_k}
[f(m^2_{N_k},M^2_Z)-f(m^2_{N_k},M^2_H)]-4M^2_Zf(m^2_{N_k},M^2_Z)
\end{equation} 
where the function $f(m^2_N,M^2)$ is defined as
\begin{equation}
f(m^2_N,M^2) \ = \ \frac{m^2_N}{m^2_N-M^2} \ln\frac{m^2_N}{M^2} \ + \
 \ln\frac{M^2}{{\mu}^2} \ -  1
\end{equation} 
Although $f$ is a $\mu$-dependent function, one can, however, show that
the final result in eq.~(4.2) does not explicitly depend on the
subtraction point $\mu$. To see that, we list the following useful
identities:
\begin{eqnarray}
\sum\limits_{k=1}^{n_G} m_{N_k} C_{iN_k} C_{jN_k} & = & 0 \ , \quad
\mbox{for} \quad i,j=1,2,\dots,2n_G\\
\sum\limits_{k=1}^{n_G} m_{N_k} B_{l_iN_k} B_{l_jN_k} & = & 0
\end{eqnarray} 
Employing eq.~(4.5), we immediately find that the $\mu$-dependence existing 
in the last term of eq.~(4.3) drops out. Eqs~(4.5) and~(4.6) also tell us
that in the limit where all heavy neutrinos are degenerated, $m^{\nu}$
approaches zero. Consequently, the smallness of the $n_G$ light neutrino
masses can be attributed to the fact that there exist $n_G$ nearly 
degenerated heavy neutrinos. Another important conclusion one can draw here is
that eq.~(4.5) does not impose any restriction on the masses of $N_i$.
Their values can be obtained in connection with the phenomenologically
constrained mixing parameters $(\xi{\xi}^{\dagger})_{{\nu}_i{\nu}_j}$~[11].
To be specific, let us consider a model with two generations only 
($n_G=2$). Then, in the leading order of $\xi$, $m^{\nu}$ takes the
simple form:
\begin{equation}
m^{\nu} \ = \ \frac{{\alpha}_W}{4\pi} \frac{M^2_H+3M^2_Z}{M^2_W}
\ln\frac{|a|}{|b|} \ \frac{1}{A} \ 
\left( \begin{array}{cc}
a^2 & ac \\
ac & c^2 \end{array} \right)
\end{equation} 
In order to obtain the mass eigenvalues, we use the freedom of rotating
the neutrino fields at the tree level (see eq.~(2.23)). Since 
$\mbox{det\hspace{0.08cm}}m^{\nu}=0$, 
this implies that one neutrino is massless,
while  the mass of the other one can be obtained by
\begin{equation}
m_{{\nu}_2} \ = \ [tr(m^{\nu}m^{\nu\dagger})]^{\frac{1}{2}}
\end{equation} 
At this loop level, the above situation is also true when one considers 
the full expression of $m^{\nu}$. Since in this two generation model
 $m^{\nu}$ is proportional to the form:
\begin{equation}
m^{\nu}_{{\nu}_i{\nu}_j} \ \propto \ C_{{\nu}_iN_2}C_{{\nu}_jN_2}
\end{equation} 
the det\hspace{0.08cm}$m^{\nu}$, 
with $m^{\nu}$ representing a tensor product of two
vectors, will always vanish. However, the above mass hierarchy is no longer
valid, when one introduces a third generation neutrino in the discussion.
In that case, the radiative neutrino mass can be written down as follows:
\begin{equation}
m^{\nu} \  = \ [g(m_{N_2})-g(m_{N_1})]m_{N_2} C_{{\nu}_iN_2}
C_{{\nu}_jN_2} + [g(m_{N_3})-g(m_{N_1})]m_{N_3} C_{{\nu}_iN_3}
C_{{\nu}_jN_3}
\end{equation} 
where $g(m_{N_i})$ are some functions that can be computed from eq.~(4.2).
So, if in eq.~(4.10), for example, we set $m_{N_1}=m_{N_3}$, then we recover
the mass spectrum of the model with $n_G=2$ discussed previously. One can
therefore conclude that the neutrino mass hierarchy will be controlled by the
heavy neutrino masses $m_{N_i}$. An extensive analysis for the more
complicated case of three generation
models will be given elsewhere [14,26].

To get an idea of some possible numerical values of $m^{\nu}$ in 
eq.~(4.7), let
\begin{equation}
a=c= 10 \ \mbox{GeV} \quad \mbox{and} \quad A=100 \ \mbox{GeV}
\end{equation} 
Then, for sufficiently small values of the perturbation parameter
$\varepsilon = (b-a)/a$, e.g. $\varepsilon \sim 10^{-4}$, we get
$m_{{\nu}_2} \sim 1-10$~eV ($M_H=100$~GeV). This neutrino mass is also
consistent with the cosmological requirement [1] that
\begin{equation}
\sum\limits_{i=1}^{n_G} m_{{\nu}_i} \stackrel{\displaystyle<}{\sim}
 \ 40 \ \mbox{eV}
\end{equation} 
Also, from eq.~(3.6) we easily derive $m_N \simeq 102$~GeV.

To further illuminate this scenario, let us numerically investigate 
the mixing couplings $(\xi{\xi}^{\dagger})_{{\nu}_i{\nu}_j}$.
They have the form:
\begin{equation}
\xi{\xi}^{\dagger} \ = \ \frac{2|a|^2}{A^2}
\left( \begin{array}{cc}
1 & 1 \\
1 & 1 \end{array} \right)
\end{equation} 
Using again the freedom of redefining the neutrino fields, 
$\xi{\xi}^{\dagger}$ can be diagonalized with eigenvalues given by
\begin{equation}
(\xi{\xi}^{\dagger})_{diag.} \ = \ \frac{4|a|^2}{A^2}
(0, \ 1)
\end{equation} 
According to~[12], the lowest upper bound on $a/A
\stackrel{\textstyle<}{\sim} 0.2$ can be obtained from the
$\tau$-neutrino mixings, whereas similar bounds coming from
${\nu}_{\mu}$, ${\nu}_e$ are more stringent ($a/A
\stackrel{\textstyle<}{\sim} 0.1$).

\setcounter{equation}{0}
\section{Phenomenological implications of large $H^0-\nu-N$ and 
$H^0-N-N$ couplings}
\indent

In this section we will discuss the phenomenological implications
of our model for the production of heavy neutral leptons and the
associated phenomena of lepton-number violation. To gauge the 
possibility of measuring such effects at the next generation
colliders ($LHC$, $SSC$, $INP$, etc.), we obtain numerical results
for production cross sections of heavy neutrinos $N_i$ -- paying
more attention on the Higgs-mediated processes. Since
there exist a variety of works studying similar phenomena~[8,10],
one may hence consider in part the discussion given here as complementary
to them. 

In order to be able to discuss the production of heavy 
neutrinos via a Higgs boson,  we summarize all $H^0$-production
cross sections at hadron and $e^+e^-$ machines in fig.~2. For numerical
estimates we have used $EHLQ$ parton distribution functions (set~2)~[22].
The subsequent decay rate of the heavy neutrinos into a specific 
final state can be described by their partial widths and branching ratios
and, as given in~[10], they are
\begin{eqnarray}
\Gamma (N \to l^{\pm}W^{\mp}) & = & \frac{{\alpha}_W}{16M^2_W}
|B_{lN}|^2 m^3_N (1+\frac{2M^2_W}{m^2_N})(1-\frac{M^2_W}{m^2_N})^2
\theta (m_N-M_W) \\
\Gamma (N \to \nu Z^0) & = & \frac{{\alpha}_W}{16M^2_W}
|C_{{\nu}N}|^2 m^3_N (1+\frac{2M^2_Z}{m^2_N})(1-\frac{M^2_Z}{m^2_N})^2
\theta (m_N-M_Z)
\end{eqnarray} 
However, in case $m_N>M_H$ another decay channel will be opened 
kinematically, given by the partial width
\begin{equation}
\Gamma (N \to \nu H^0) = \frac{{\alpha}_W}{16M^2_W} 
|C_{{\nu}N}|^2 m^3_N (1-\frac{M^2_H}{m^2_N})^2 \theta (m_N-M_H)
\end{equation} 
The branching ratio for a situation where $m_N \gg M_W,M_Z,M_H$ and 
$C_{{\nu}N} \simeq B_{lN}$ is
\begin{equation}
\sum\limits_{i=1}^{n_G} Br( N \to {\nu}_i Z^0) =
\sum\limits_{i=1}^{n_G} Br( N \to l^+_iW^-) =
\sum\limits_{i=1}^{n_G} Br( N \to {\nu}_i H^0) = \frac{1}{4}
\end{equation} 
Instead, if $m_N \stackrel{\textstyle<}{\sim} M_H$ the above ratio for the 
different decay modes increases up to $1/3$.

The Higgs-mediated processes at high energies can produce heavy neutrinos
$N_i$  through the $H^0-N-N$ and $H^0-\nu-N$ couplings. The Majorana nature
of the heavy neutrinos in the first class of reactions (i.e.
$e^+e^-, \ pp \to H^{0\ast} \to NN$) may be proved by detecting like-sign
dilepton pairs associated with jets with no missing $p_T$~[23]. The
second class of processes (i.e. $e^+e^-,\ pp \to H^{0\ast} \to N\nu$) will be
more problematic. Nevertheless, if the standard background contributed
to Higgs decays could be theoretically  removed, 
the detection of neutral leptons 
via a flavour-nonconserved $H^0-N-\nu$ coupling would be favourably
interpreted as an indication of heavy Majorana neutrino events. Another
feature of these processes is the very large missing $p_T$ at the
Higgs-resonance line. Since we are interested in the production rate
of $N_i$ via a heavy Higgs boson, let us, for completeness, quote
the main partial decay widths.
\begin{eqnarray}
\Gamma (H^0 \to {\nu}_i N_j) & = & \frac{{\alpha}_WM_H}{8}
|C_{{\nu}_iN_j}|^2 \frac{m^2_{N_j}}{M^2_W} (1-\frac{m^2_{N_j}}{M^2_H})
\theta (M_H-m_{N_j}) \\
\Gamma (H^0 \to N_i N_j) & = & {\delta}_{ij}\frac{{\alpha}_WM_H}{4}
(ReC_{N_iN_j})^2 \frac{m^2_{N_i}}{M^2_W} \Big[
1-4\frac{m^2_{N_i}}{M^2_H} \Big]^{\frac{3}{2}} \theta (M_H-m_{N_i}-
m_{N_j}) \nonumber\\
 & & 
\end{eqnarray} 
In eq.~(5.6) we have assumed that $N_i$ are nearly degenerated as dictated
by this model. In addition, the phenomenology of the Higgs boson
in the $SM$ is studied rather extensively 
at present~[24,25]. Fig.~3 shows that for $m_N \simeq 100$~GeV the
channel $H^0 \to {\nu}N$ could be the most dominant decay mode in the
intermediate Higgs mass range (i.e $100\leq M_H \leq 150$~GeV). Such
events are characterized by a very large missing transverse momentum, 
a fact that can probably be exploited to reduce efficiently the
contributing background
\footnote[1]{Similar technics relying on particular kinematical cuts
are used, for example, for the reconstruction of the production
of supersymmetric particles at hadron colliders~[27]. In any case, the
viability of such heavy neutrino signals from the background will be
investigated in~[14].}.
 In particular, when $B_{lN}$ 
is suppressed but $C_{{\nu}N}$ not~[10], the reaction
$e^+e^- \to H^{0\ast} \to \nu N X$ will become the most significant
production mechanism for 100~GeV neutrinos with a cross section
value ${\sigma}_{tot} \simeq 0.3 \ pb$ at {\em c.m.s.} energies
$\sqrt{s_{tot}} = 1-2$~TeV. 

The branching ratios for the decay of the Higgs scalar into 
the modes $NN$ and $\nu N$
neutrinos are generally given in figs~4 and~5. More precisely, for 
$M_H \simeq 3m_N$, we get a maximum value for 
$Br(H\to NN) \simeq 1.6\ 10^{-3}$ and for $M_H \simeq 1.5m_N$ we
have a maximum value for $Br(H \to N{\nu}) \simeq 3 \ 10^{-2}$.
Illustratively, we mention that for
$m_N=150$~GeV, $M_H \simeq 450$~GeV and $a/A=0.2$ we expect
about 480 lepton-violating events per year at $LHC$ 
($\sqrt{s_{tot}} =16$~TeV) assuming the standard high luminosity
${\cal L}=4 \ 10^5 pb^{-1}/$year. The corresponding rate at $SSC$
($\sqrt{s_{tot}} = 40$~TeV, ${\cal L} = 10^4\ pb^{-1}/$yr) is relative
small, i.e. about 80 equal-sign dileptons a year.

To have a complete picture, in fig.~6 we give the cross-section values
of the $W$-mediated tree process, relevant for the production
of dilepton signals with no missing $p_T$ at $pp$~machines, 
as a function of $m_N$. The production cross section is evaluated by
\begin{equation}
\sigma(s_{tot}) = 2 \int dx \int dy [f^p_{\bar{u}}(x,Q^2)
f^p_d(y,Q^2)+f^p_{\bar{c}}(x,Q^2)f^p_s(y,Q^2)] \hat{\sigma}(\hat{s})
\end{equation} 
where $f^p$'s are parton distribution functions~[22] at $Q^2=\hat{s}=
xys_{tot}$ and $x$, $y$ are usual kinematical variables restricted
to the intervals
\begin{equation}
\frac{m^2_N}{s_{tot}} \leq x \leq 1  \  \ , \quad
\frac{m^2_N}{xs_{tot}} \leq y \leq 1 
\end{equation} 
Moreover, the subprocess cross section $\hat{\sigma}$ is given by
\begin{equation}
\hat{\sigma} (\hat{s}) = \frac{{\pi}{\alpha}^2_W}{72
{\hat{s}}^2(\hat{s}-M^2_W)^2} |B_{lN}|^2
(\hat{s}-m^2_N)^2(2\hat{s}+m^2_N)
\end{equation} 
Now, we can easily find that for unsuppressed $W-l-N$ couplings (e.g.
$B_{lN} \simeq C_{{\nu}N}$), this scattering process gives a large
amount of lepton-number-violating signals. Specifically, for $m_N=150$~GeV
one could expect up to 10000(1000) events with no missing $p_T$ at
$LHC(SSC)$. As a consequence, high energy colliders will 
successfully explore such heavy neutrino scenarios for the first time
and may indirectly lead us to some important clues concerning 
the nature of ordinary neutrinos. 

\setcounter{equation}{0}
\section{Conclusions}
\indent 

We have presented a new radiative mechanism of generating small
neutrino masses, in the simplest model which predicts heavy Majorana
neutrinos, i.e. the $SM$ with one right-handed neutrino per family.
This mechanism based on large intergeneration mixings of
"democratic" type (${\xi}_{{\nu}N} \sim 0.1$) naturally provides very light
neutrinos with $m_{\nu} \stackrel{\textstyle<}{\sim} 10$~eV at the first
electroweak loop level and nearly degenerated heavy neutrinos $N_i$ with
$m_N$ being in the 100~GeV range. On the contrary, from the ordinary "see-saw"
mechanism and taking $m_{D_i} \sim m_{l_i}$, 
one derives the values $\sim 10^{-7}$~eV, 
$10^{-2}$~eV, 10~eV for $m_{{\nu}_e}$, $m_{{\nu}_{\mu}}$,
$m_{{\nu}_{\tau}}$, respectively, with a heavy neutrino
mass  $m_N \sim 10^9$~GeV. The common feature of
such cosmologically consistent scenarios
proposed by many authors~[7] is that they generally require 
an extremely large Majorana scale ($m_N \sim m_M \sim 10^8 - 10^{12}$~GeV)
and are mostly associated 
with invisible axions. However, our 
suggested scheme is a rather interesting and viable possibility which
implies heavy neutral leptons with a rather low mass scale 
$m_M \sim 100$~GeV.

In section~5, we have discussed the phenomenological implications
of the model under consideration for the production of heavy neutrinos
$N_i$ and the associated lepton-number violating signatures. Paying special
attention on the Higgs sector of the model, we have found numerically
that Higgs bosons may predominantly decay into heavy Majorana neutrinos
with $m_N \simeq 100$~GeV for $100 \leq M_H \leq 150$~GeV. Furthermore,
for heavier neutrino masses the Higgs-mediated processes at $LHC$ can give 
rise to a sufficiently large number of like-sign dileptons with no missing
$p_T$ of the order of $10^3$ events a year. 
For comparison, we have also numerically
evaluated the tree-level $W$-exchange process,
$ pp \to W^{-\ast} \to l^-NX$ (see also fig.~6).
This provides an event rate of lepton-number-violating signals up to 
one hundred times larger  than the $H^0$-exchange processes,
when an unsuppressed $W-l-N$ coupling is assumed.

A large $H-\nu-N$ coupling may also have important consequences 
on obtaining large lepton-flavour-nonconservation
phenomena in $H^0$~decays~[26]. Such signals
characterized by no missing~$p_T$ and no hard jet events can easily
be reconstructed experimentally and may be particularly useful for a clear
observation of an intermediate Higgs boson at hadron colliders.
All these new theoretical aspects of the model discussed in this work 
may lead us to further theoretical considerations in future
and could therefore constitute an additional motivation for us to 
explore new physics, which may hopefully open up another
possibility of investigating the nature of the neutrino particles.

\section*{Acknowledgements}
\indent

I am grateful to A. Datta for useful discussions and comments.
I also thank E.~A.~Paschos, C.~T.~Hill, J.~G.~K\"orner, 
 K.~Schilcher and Y.~L.~Wu for helpful hints and conversations, and
B.~K\"onig for a critical reading of the manuscript.
This work has been supported by a grant from the Postdoctoral 
Graduate College of Germany.

\newpage
\section*{Figure Captions}
\newcounter{fig}
\begin{list}{\bf\rm Fig. \arabic{fig}:}{\usecounter{fig}
\labelwidth1.6cm \leftmargin2.5cm \labelsep0.4cm \itemsep0ex plus0.2ex \sl}

\item Graphs relevant for the radiatively induced neutrino mass matrix
$m^{\nu}$ in the Feynman gauge.

\item Production cross sections of the Higgs boson as function of its mass
at different collision machines: $SSC$~(dashed line), $LHC$~(solid line),
2-TeV $e^+e^-$ collider~(dashed-dotted line), 1-TeV $e^+e^-$ 
collider~(dotted line).

\item The behaviour of the ratio $R=\Gamma (H^0 \to \nu N)/\Gamma (H^0
\to b \bar{b})$ for A=100~GeV and different values of $a/A$:
0.2~(dotted line), 0.1~(dashed line), 0.05~(solid line).

\item The branching ratio of Higgs decays into two heavy neutrinos.
We have set $a/A=0.2$ and $A=150$~GeV (solid line), 200~GeV~(dashed
line), 250~GeV~(dotted line).

\item The branching ratio of Higgs decays into $\nu N$ for 
$A=200$~GeV and $a/A=0.1$~(solid line), $A=200$~GeV and 
$a/A=0.2$~(dashed line), $A=400$~GeV and $a/A=0.1$~(dotted line),
$A=400$~GeV and $a/A=0.2$~(dashed-dotted line).

\item The total cross section of the reaction $pp \to W^{-\ast} \to
l^- N X$ at $SSC$~(solid line) and at $LHC$~(dashed line).

\end{list}

\end{document}